\documentstyle[11pt,paspconf,epsf]{article}
\def \lya {Ly$\alpha$}

\def \ha {H$\alpha$}
\def \hb {H$\beta$}
\def \oIII {[O~III]~$\lambda\lambda 4959, 5007$}
\def \kms {~km~s$^{-1}$}
\def \cmII {cm$^{-2}$}
\def \cmIII {cm$^{-3}$}

\def \zw {I~Zw~1}
\def\lesssim{\mathrel{\hbox{\rlap{\hbox{\lower4pt\hbox{$\sim$}}}\hbox{$<$}}}}
\def\gtrsim{\mathrel{\hbox{\rlap{\hbox{\lower4pt\hbox{$\sim$}}}\hbox{$>$}}}}
\begin{document}

\title{The UV Properties of the Narrow Line Quasar I~Zwicky~1}
\author {Ari Laor}
\affil{Physics Department, Technion, Haifa 32000, Israel}

\begin{abstract}

\zw\ is the prototype narrow line quasar. This type of AGNs show a number of 
common peculiar characteristics whose physical origin is not yet well understood.
Here, I review the results of a detailed study by Laor et al. (1997b) of the UV 
emission properties of \zw, based on a high quality {\em HST} FOS spectrum. The 
main results are: 1) The Mg~II and the Al~III doublets are resolved, indicating 
optically thick thermalized emission, as expected for a standard BLR. 
2) Lines from ions of increasing ionization level show increasing excess 
blue wing flux, and an increasing line peak velocity shift, suggesting an 
out-flowing component of the BLR, visible only in the approaching direction,
where the ionization level increases with velocity. 3) A weak 
associated UV absorption system is detected in N~V, C~IV, Si~IV and \lya, which 
is probably part of the outflow suggested above. 4) The C III]~$\lambda 1909$ 
line indicates a BLR density about an order of magnitude larger than commonly 
observed in quasars. An additional denser component is implied by the 
Al~III~$\lambda 1857$ doublet. 5) Prominent Fe~II and Fe~III emission is seen, 
including components which may indicate significant \lya\ pumping of the 8-10 eV 
levels of Fe~II. Further studies suggest that some of 
the above properties are typical of narrow line AGNs. 

\end{abstract}


\section{Introduction}

\zw\ is the prototype narrow line Seyfert 1 galaxy (NLS1, although
it luminosity, $M_V=-23.8$ for $H_0=50$\kms~Mpc$^{-1}$, qualifies
it as a low luminosity quasar). Its
optical spectrum reveals narrow emission
lines and strong Fe~II emission (Phillips 1976, 1978; 
Oke \& Lauer 1979). The NLS1s also tend
to have weak \oIII\ emission, asymmetric \hb\ profiles 
(Boroson \& Green 1992), 
steep soft X-ray spectra (Laor et al. 1994, 1997a; Boller, 
Brandt \& Fink 1996), rapid X-ray variability (Fiore et al. 1998),
and possibly strong IR emission (e.g. Lipari 1994). The physical origin
for this set of peculiar emission properties is not well understood yet.
In this contribution I review the main results of a detailed study of a
high quality UV emission line spectrum of \zw\ obtained with the {\em HST} FOS
(for further details see Laor et al. 1997b). This study provides additional clues
to the physical processes responsible for the peculiar properties of NLS1s.

\section{Observations} 

We obtained high S/N spectra (50$-120$ per resolution element) 
of \zw\ with 
the three high resolution (R=1300) gratings of the FOS from
1150~\AA\ to 3280~\AA. This S/N is comparable
to the highest S/N yet obtained with HST for bright AGNs. 
We also obtained a complimentary high resolution ground based
spectrum which covers the range from 3183~\AA\ to 4074~\AA.

\begin{figure}
\plotone{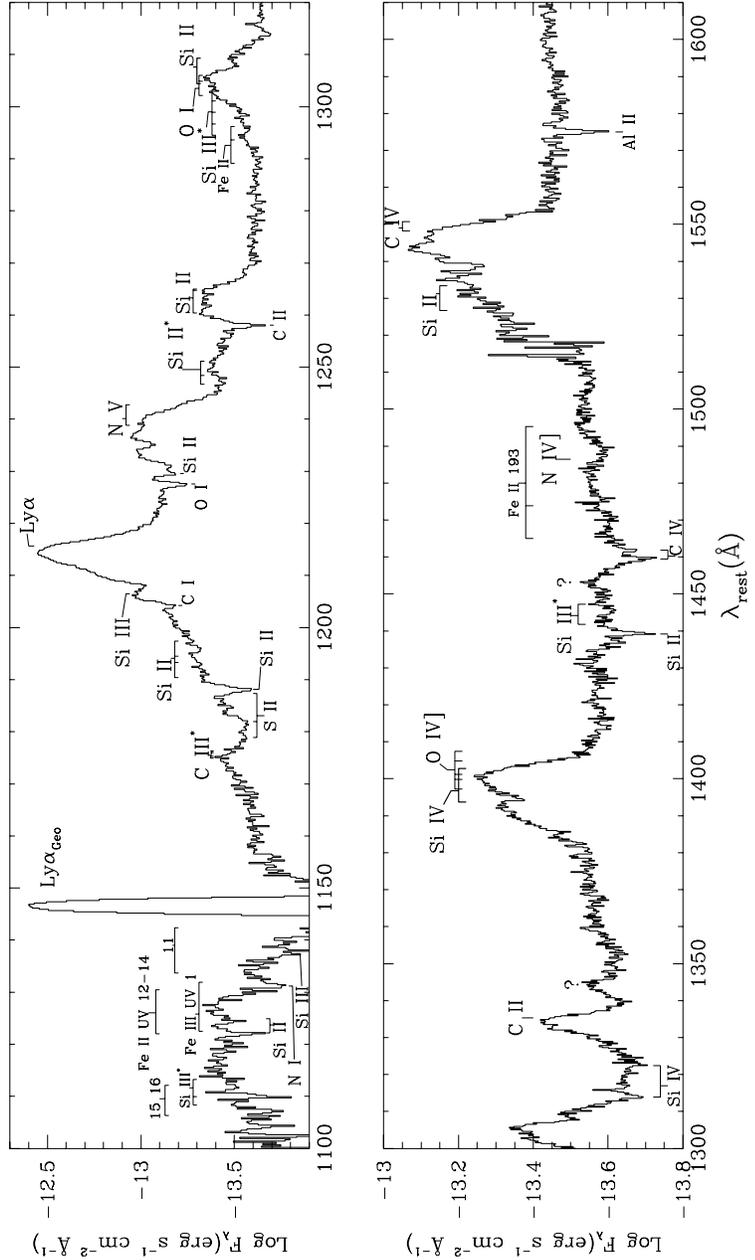}
\caption{Tentative identifications of most of the emission features in the HST FOS 
spectrum of \zw. The lines indicate the expected wavelength for $z=0.0608$.
Only a small number of the Fe~II multiplets which may be present
were marked. Line designations below the spectrum refer to Galactic
absorption features.} \label{fig-1}
\end{figure}
\begin{figure}
\plotone{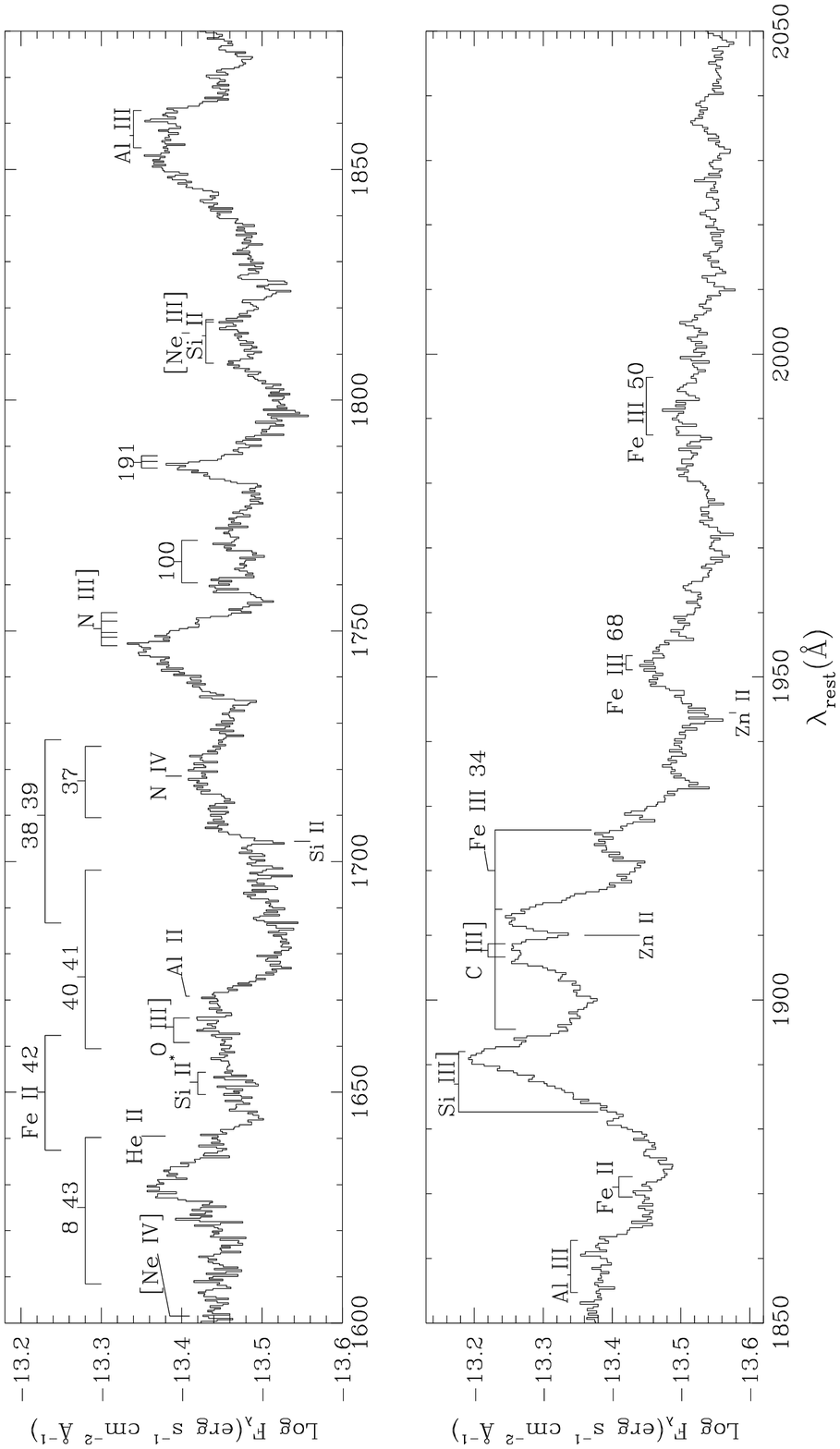}
\caption{As in figure 1.} \label{fig-2}
\end{figure}

Figures 1 and 2 present an expanded view of the 1100\AA\ to 2000\AA\
spectrum of \zw, where most of the strong UV lines are present. The spectrum 
was corrected for the effects of Galactic extinction of E(B$-$V)=0.1, and
converted to rest wavelength 
assuming $z=0.0608$, the redshifts of \ha\ and \hb\
(e.g. Phillips 1976).
The rest frame positions of various lines are marked above the spectrum.
Lines marked below the spectrum designate various absorption lines
originating in the Galactic interstellar medium (ISM). All ISM lines
detected here are typically seen in FOS spectra of quasars. 
We made no attempt to identify the many Fe~II emission blends 
 present. The various Fe~II multiplets designated in Fig.2 
are taken from Moore (1952), and serve
only as an illustration of possible Fe~II features. The number of possible
Fe~II multiplets is extremely large, and a reliable identification requires
detailed theoretical calculations.
The complete spectrum, and the measured emission line parameters are provided
in Laor et al. (1997b).
 
\section{Line Diagnostics} 

\subsection{C}

The [C~III]~$\lambda 1907$ line may be present in \zw. This line
is produced by a forbidden transition from the 
$^3\!P_2$ term in the $2s2p$ configuration of C~III down to the
ground level, with a critical density of $n_c=5\times 10^5$~\cmIII.
The $^3\!P_2$ term lies 0.007~eV 
above the $^3\!P_1$ term which produces the well known C~III]$~\lambda 1909$ 
line, where $n_c\sim 5\times 10^9$~\cmIII.
Thus, the flux ratio 
$R_{\rm C~III}\equiv$[C~III]~$\lambda 1907/$C~III]$~\lambda 1909$ 
provides a temperature independent density diagnostic (e.g. Osterbrock 1989,
Fig. 5.5). The observed $R_{\rm C~III}\lesssim 0.1$ implies
that some of the C III emission originates in gas with 
an electron density $n_e\lesssim 5\times 10^5$~\cmIII. 

The C~III]$~\lambda 1909$ line is very weak in \zw. Both its EW of 
$\sim 2.5$~\AA, and the C~III]$~\lambda 1909/$\lya=0.023 flux ratio
are a factor of 5-10 smaller than typically observed
(e.g. Laor et al. 1995; Wills et al. 1995).
Since C~III]$~\lambda 1909$ is thermalized at 
$n_e\gtrsim 5\times 10^9$~\cmIII,
the suppression of C~III]$~\lambda 1909$ may imply that the typical BLR density 
in \zw\ is about an order of 
magnitude larger than indicated by C~III]$~\lambda 1909$ in typical AGNs, 
i.e. $n_e\sim 10^{11}$~\cmIII\ instead of 
$n_e\sim 10^{10}$~\cmIII. Additional evidence for this interpretation is
provided by the Si~III]$~\lambda 1892$ line discussed below.

\subsection{Mg}

The Mg~II ion survives mostly  
in the partially ionized H region, where photons above 13.6~eV are
blocked. The Mg~II~$\lambda 2798$ doublet is therefore a diagnostic of the
conditions in the extended partially ionized layers inside the BLR
clouds. The flux ratio of the Mg~II 2796.35~\AA/2803.53~\AA\ components 
indicates the optical depth in the line. 

 The 2803.53~\AA, 2796.35~\AA\ lines correspond to
transitions from the $3p\,^2\!P_{1/2}$ and $3p\,^2\!P_{3/2}$
levels to the ground $3s\,^2\!S_{1/2}$ level. 
Collisional de-excitation 
dominates when $\tau_0 n_e>A/q$, 
where $\tau_0$ is the line center optical depth for resonance scattering, 
i.e when $\tau_0 n_e>7.1\times 10^{14}$~cm$^{-3}$ and 
$3.6\times 10^{14}$~cm$^{-3}$ for the $3p\,^2\!P_{3/2}$ and $3p\,^2\!P_{1/2}$
levels. For $\tau_0 n_e>7.1\times 10^{14}$~cm$^{-3}$ 
the Mg~II 2796.35~\AA/2803.53~\AA\ doublet ratio is thermalized, 
i.e. it changes from 2/1 to 1/1 (see Hamann et al. 1995 for results
on other analogous UV doublets). Thus, the observed 1.2/1 ratio implies
that most of the Mg~II emission 
is produced in a region where 
$\tau_0 n_e\gtrsim 10^{15}$~cm$^{-3}$. 

As described in Laor et al. (1997b), the value of $\tau_0 n_e$ provides
a direct measure of the distance of the BLR clouds from the central
ionizing continuum source. The
line center optical depth is $\tau_0=\sigma_{\nu_0}\Sigma_{\rm Mg~II}$,
where $\sigma_{\nu_0}=1.47\times 10^{-12}$~cm$^2$ is the total 
Mg~II doublet line center absorption cross section (for $T=10^4$~K),
and $\Sigma_{\rm Mg~II}$ is the column density of Mg~II ions.
The partially ionized H region, where Mg~II survives, has a column 
density about
10-30 times larger than the column density of the highly ionized H region, 
given
by $\Sigma_{\rm H~II}=10^{23}U$~\cmII, where $U$ is the ionization
parameter ($\equiv n_{\gamma}/n_e$, the photon/electron density). 
Balmer continuum absorption may limit the depth of the Mg~II doublet 
emitting layer, but this absorption will not affect our conclusion. 
Using the Mg/H solar abundance ratio of $3.8\times 10^{-5}$, and assuming 
most Mg is Mg~II, we get 
$\Sigma_{\rm Mg~II}\simeq 4\times 10^{19}U$~\cmII, or
$\tau_0=6\times 10^7U$, assuming the gas is ionization, 
rather than matter bounded. The Mg~II doublet is therefore thermalized
when $6\times 10^7Un_e\gtrsim 10^{15}$, or $n_{\gamma}\gtrsim 10^7$~\cmIII,
which is consistent with the standard estimate of 
$n_{\gamma}\sim 2\times 10^9$~\cmIII\ in the BLR.

\subsection{Al}

The components of the Al~III~$\lambda 1857$ doublet
are widely spaced ($\sim 1300$\kms) and are therefore clearly
resolved in \zw. Al~III is created by photons above 18.829~eV, and 
destroyed by photons above 28.448~eV, thus unlike Mg~II, it exists in 
the highly ionized H region only, and the observed Al~III doublet ratio
provides constraints on the BLR H~II region, as discussed below.

The template fit suggests a 1.25/1 ratio for the 
1854.71~\AA/1862.79~\AA\ doublet line. 
 Baldwin et al. (1996) found a similar 1/1 flux ratio in
three objects where the Al~III doublet was clearly detected.

The Al~III and Mg~II ions are both Na~I like.
Using the same reasoning described for Mg~II we find that the 
Al~III~$\lambda 1857$ doublet is thermalized
for $n_{\gamma}\gtrsim 4\times 10^9$~\cmIII, and the 
observed 1.25/1  doublet ratio indicates that
it is formed in clouds at, or inside the `standard'
BLR radius of $0.1L_{46}^{1/2}$~pc, where 
$n_{\gamma}\gtrsim 2\times 10^9$~\cmIII. 

The relatively large Al~III~$\lambda 1857$ EW (4.4~\AA) requires a
BLR component where Al~III is the dominant ionization state of Al,
which occurs for $U\sim 10^{-2}-10^{-3}$ (e.g. Baldwin et al.
1996, Fig.7a). 
Since $n_{\gamma}\gtrsim 4\times 10^9$~\cmIII, the above range
in $U$ implies 
$n_e\gtrsim 4\times 10^{11}-4\times 10^{12}$~\cmIII, which is comparable,
or denser, than the component responsible for the C~III]~$\lambda 1909$ 
emission.

\subsection{Si}

The Si~II 1260.42~\AA, 1264.74~\AA, 1265.00~\AA, 1304.37~\AA, 
1309.28~\AA, 1808.01~\AA, 1816.93~\AA, and 1817.45~\AA\ lines are
clearly detected. The flux of the Si~II $\lambda 1194$ and $\lambda 2335$ 
blends is well constrained. This wealth of data provides a 
unique opportunity to explore the Si~II line formation mechanisms
(e.g. Baldwin et al. 1996, Appendix C). 
 
The outer shell electronic configuration of the Si~III ion is
analogous to that of C~III. The analogous line to the 
[C~III]~$\lambda 1907$ line is the [Si~III]~$\lambda 1883$
line, which can also be used as a density diagnostic for the
Si~III gas. Weak [Si III]~$\lambda 1883$ emission may be present 
at 1881.1~\AA. As for the C~III ion, the 
$R_{\rm Si~III}\equiv$~[Si~III]~$\lambda 1883/$Si~III]~$\lambda 1892$ 
flux ratio is a robust density
indicator. The observed 
$R_{\rm Si~III}\lesssim 0.1$ implies that some of the 
Si~III resides in gas with
$n_e\sim 5\times 10^5$~\cmIII, i.e. the same conditions
implied by $R_{\rm C~III}$, as expected since the spatial 
distribution of both ions in photoionized gas should largely overlap. 

The Si~III]~$\lambda 1892$/C~III]~$\lambda 1909$ flux ratio of $\sim 3.5$
is significantly larger than the typical ratio
of $\sim 0.3\pm 0.1$ observed in quasars (Laor et al. 1995). This
high ratio results from the factor of 5-10 suppression in the 
C~III]~$\lambda 1909$ flux, rather than significant
enhancement of Si~III]~$\lambda 1892$. The most likely interpretation 
for this high ratio is a relatively dense BLR.
The critical density for C~III]~$\lambda 1909$,  
$n_c\sim 5\times 10^9$~\cmIII, is significantly smaller than 
for Si~III]~$\lambda 1892$, where $n_c=1.1\times 10^{11}$~\cmIII.
Thus, if the BLR density is $10^{11}$~\cmIII, rather than $10^{10}$~\cmIII,
C~III]~$\lambda 1909$ would be suppressed by a factor of $\sim 10$,
while Si~III]~$\lambda 1892$ would not be affected, as observed.

\subsection{Fe}

The likely presence of Fe~II emission well below 2000~\AA, and
possibly even at $\sim 1110$-1130~\AA, indicates that Fe~II is
excited by processes other than just collisions, as the electron
temperature in the Fe~II region is most likely 
far too low for significant
collisional excitation of levels $\gtrsim 10$~eV above the ground state. 
One such process,
suggested by Penston (1987), is resonance absorption of \lya\
photons by Fe~II. Johansson \& Jordan (1984) found \lya\ 
resonance absorption to be significant in various stellar systems,
and identified the Fe~II~$\lambda 1294$,  
Fe~II~UV~191~$\lambda 1787$, and the Fe~II~$\lambda 1871$ multiplets
as the signatures of such a process. 
The Fe~II~UV~191 multiplet is very
prominent in \zw, and weak emission blends are clearly apparent
at 1294~\AA, and 1870~\AA, suggesting that resonance absorption
of \lya\ photons may be a significant excitation mechanism
for Fe~II in \zw\ as well. Resonance scattering of continuum photons, 
and Fe~II-Fe~II line flouresence could also be 
a significant process for populating high lying Fe~II levels
(e.g. Netzer \& Wills 1983; Wills, Netzer \& Wills 1985).
Clearly, the Fe~II rich spectrum of \zw\ should serve as a 
valuable tool for future studies of Fe~II emission in AGNs.

A number of Fe~III multiplets are clearly present.
The Fe~III~UV~34 multiplet, clearly identified here, was
first discovered by Hartig \& Baldwin (1986) in a 
broad absorption line quasar. An emission feature near $\sim 2070$~\AA\
is commonly seen in quasars (e.g. Wills et al. 1980), 
and was identified as a likely Fe~II blend. Baldwin et al. (1996)
suggested that this feature is due to Fe~III~UV~48 emission. Here the
three Fe~III~UV~48 components at 2062.2~\AA, 2068.9~\AA, and 2079.65~\AA\
are clearly resolved, which verifies the identity of this feature.
The Fe~III~UV~47~$\lambda\lambda 2419.3, 2438.9$ doublet is most likely 
present as well.
Other Fe~III blends, such as Fe~III~UV~50 and Fe~III~UV~68,
and in particular the resonance Fe~III~UV~1 at 1122-1130~\AA\ 
may also be present, but these identifications cannot be verified
here since these blends are not clearly resolved.

The relative flux ratio of the multiplet components can provide
an important diagnostic for the optical depth and excitation mechanism 
of Fe~III.

\subsection{Line Profiles and Velocity Shifts}

Optical spectra of \zw\ revealed two velocity systems. 
The first system is at $z=0.0608$ for the low ionization forbidden lines 
and Balmer lines, including \ha, \hb, optical Fe~II multiplets, [Ca~II],
[S~II], [N~II], He~I, Na~I, and [O~I]. The second 
system is at $z=0.0587$ (i.e. blueshifted by $\sim 630$~\kms) for the 
higher ionization forbidden lines,
[O~III] and [Ne~III] (Phillips 1976, Oke \& Lauer 1979).

We find the same trend in the UV. The low 
ionization permitted lines, including O~I, 
C~II, C~II], N~II], Al~II, 
Al~II], the various Si~II multiplets,
and Fe~II~191 are all blueshifted by 
$\lesssim 200$\kms\ with respect to $z=0.0608$. 
 Higher ionization lines, including N~III], Al~III,
Si~III] and Si~IV
are blueshifted by $\sim 300-500$\kms, while the
highest ionization lines, C~IV and
N~V, are blueshifted by $\sim 900$\kms.
The highest blueshift, $\sim 2000$\kms, 
is displayed by He~II~$\lambda 1640.4$.
This trend of increasing blueshift with increasing ionization level is 
observed in most quasars 
(e.g. Gaskell 1982; Wilkes 1984, 1986; Espey et al. 1989). However, the 
typical amplitude of the
high ionization line blueshift, such as C~IV and N~V,
 is only $\sim 200-300$\kms,
and for He~II it is $\sim 500$\kms\ (Tytler \& Fan 1992; Laor et al. 1995), 
i.e. about four times lower than found in \zw.

The low ionization line profiles are mostly consistent with the rather
symmetric \ha\ profile.
With increasing blueshift the lines get progressively broader and develop 
a progressively stronger blue excess asymmetry (fig.3 in Laor et al. 1997b). 
  
\section{UV absorption}

Associated absorption is present in \lya, N~V, C~IV and Si~IV. 
The absorption is
blueshifted by $\sim 1870$\kms\ relative to the $z=0.0608$ frame, it has
a FWHM$\sim 300$\kms, and an absorption EW of $\sim 0.25$~\AA\ for most
lines. 
Since the spectral resolution of the
FOS is $\sim 230$\kms, the absorption system may not be truly 
resolved. 

The presence of high ionization lines, and lack of low ionization lines
in absorption
suggests the absorber is associated with \zw, rather
than with an intervening system unrelated to \zw\ (see e.g. Hamann 1997).
Most of the constraints below are independent of the exact location of the
absorber.

\subsection{Implied constraints}

The column density associated with the observed absorption is a 
function of the absorber optical depth (assuming a full coverage). 
The optical depth can be deduced
from the ratio of EW of the two doublet components. The observed ratio 
in N~V is $\sim 1.4$, but it is
rather uncertain, and thus the absorption optical depth 
remains uncertain. 

The absorbing columns are in the range
$N_{\rm H~I}=4.6\times 10^{13}-4.3\times 10^{14}$~cm$^{-2}$,
$N_{\rm C~IV}=1.2\times 10^{14}-1.4\times 10^{18}$~cm$^{-2}$, and
$N_{\rm N~V}=2.3\times 10^{14}-3.3\times 10^{18}$~cm$^{-2}$. 

To infer the total H column density from the H~I column density 
one needs to know
the H ionization state. The 
H~II/H~I fraction in the absorbing gas is related to the ionization
parameter through 
$ N_{\rm H~II}/N_{\rm H~I}\simeq 2\times 10^5 U $ 
(e.g. Netzer 1990). Thus, the upper limit on the H column density is
$ N_{\rm H}\simeq N_{\rm H~II}=(0.2-6)\times 10^{20}U~{\rm cm}^{-2}$.
The presence of N~V and C~IV ions suggests that
the absorber has $U\sim 0.01-1$, thus the H column density 
upper limit is probably
in the range $10^{17}\lesssim N_{\rm H} \lesssim 6\times 10^{20}$.
If the absorber metal abundance is solar then the above $N_{\rm H}$ implies
metal columns upper limits of
$3.6\times 10^{13}\lesssim N_{\rm C} \lesssim 2\times 10^{17}$ and
$10^{13}\lesssim N_{\rm N} \lesssim 6.7\times 10^{16}$. These values
overlap with the directly determined constraints on the 
C~IV and N~V columns and suggests that
significant fractions of C and N may indeed be in the form of C~IV and N~V,
consistent with the assumption of $U\sim 0.01-1$. Detection of 
O~VI~$\lambda\lambda 1031.93, 1037.63$ absorption would allow a much tighter 
constraint on the absorber's $U$.

\subsection{UV and optical emission}

The UV absorber blueshift of $\sim 1870$\kms\ is remarkably close to the
He~II~$\lambda 1640$ emission line blueshift of $\sim 1990$\kms, suggesting that the
strongly blueshifted He~II emission line peak 
may originate in the same outflowing gas which produces the
UV absorption lines. In order to emit the observed He~II~$\lambda 1640$ flux the
absorber must absorb most photons just above the He~II bound-free edge
at 4~Rydberg. If the column density of the UV absorber is close to the upper limit, and
its covering factor is close to unity, it may contribute
significantly to the observed He~II~$\lambda 1640$ emission.

\section{Discussion} 

The strong optical Fe~II emission of \zw, its weak [O~III] emission, 
strong IR emission, and ``red'' UV continuum are all typical properties of
low ionization broad absorption line quasars (BALQSOs; Weymann et al. 1991, 
Boroson \& Meyers 1992; Sprayberry \& Foltz 1992). The presence of weak UV 
absorption at a blueshift of $\sim 2000$\kms\ 
in \zw\ suggests it may be a ``failed'' low ionization BALQSO, i.e.
a BALQSO where our line of sight just grazes the outflowing high ionization
wind, and misses the low ionization outflow, as suggested by Turnshek et al.
(1994) in the case of PG~0043+039. 

Is the red continuum, excess blue flux in the
high ionization lines, associated absorption, and dense BLR, unique to \zw,
or are they typical UV properties of narrow line quasars? 
No complete UV study of narrow line AGNs is available to answer these
questions, but some hints may be obtained from existing 
observations. 

Baldwin et al. (1996) studied the emission line properties of a heterogeneous
sample of seven $z\sim 2$ 
quasars with a range of emission line properties. Some of their objects have 
strong Fe~II, Fe~III, and Al~III
emission as observed in \zw. Baldwin et al. analyzed in
detail the UV spectrum of Q$0207-398$, where a high S/N ratio was available,
and found excess emission in the blue wing of the high ionization UV lines,
which they interpreted as a high ionization outflowing component in the BLR,
although there was no direct evidence (through absorption) for such a component,
as observed in \zw. 

It is also interesting to note that Boroson \& Green found that
strong optical Fe~II emitting quasars, which tend to have narrow \hb, also 
tend to have blue excess flux in \hb, and Boroson \& Meyers found that low
ionization BALQSOs, which are generally \zw\ like, have blue excess flux in \ha.
It is not known, however, whether this
property extends to the UV lines as well.

If the UV absorption system in \zw\ is indeed producing the blueshifted
components of the high ionization lines, then the absence of a corresponding
redshifted component indicates that the far side of this
outflow has to be obscured. This may either be due to an extended highly
optically thick gas, such as an accretion disk, or it may result from
absorption within each cloud, if the
clouds column density is large enough (see Ferland et al. 1992 for 
discussion).

\zw\ has particularly weak C~III]~$\lambda~1909$ emission 
but normal Si~III]~$\lambda~1892$ emission, which
we interpret here as evidence for a relatively dense ($\sim 10^{11}$~\cmIII)
BLR, at least for the region which produces the low ionization lines.
 Baldwin et al. (1988) noted the weakness of C~III]~$\lambda~1909$ 
in four other quasars with
narrow UV lines, and this trend is also apparent in Baldwin et al. (1996)
quasars. The most extreme case is of H$0335-336$ where essentially
no C~III]~$\lambda~1909$ emission was observed (Hartig \& Baldwin 1986). 
This quasar has very narrow lines, and is also a low ionization BALQSO.

Independent evidence for a dense BLR with 
$n_e\gtrsim 10^{11}$~\cmIII\ is provided by the significant EW
and thermalized doublet ratio of the Al~III~$\lambda 1857$ doublet, 
and by the significant EW of C~III$^*~\lambda 1176$. 
Another indication for a dense BLR in \zw\ is provided by
its optical spectrum. The Na~I~$\lambda\lambda 5890, 5896$ emission
line is rather
strong (Oke \& Lauer 1979; Phillips 1976), which Thompson (1991) and
Korista et al. (1997, their Figure 3g) find can only be produced for 
$n_e\gtrsim 10^{11}$~\cmIII, and a large
column density (required to shield neutral Na~I from ionizing 
radiation at $E>5.14$~eV).

It thus appears that some of the properties of \zw, in particular the 
relatively weak C~III]~$\lambda~1909$, and the blueshifted excess emission
in the high ionization lines, may be common in narrow line quasars.
A more systematic study of the UV emission of narrow line
AGNs is required to establish their typical emission line properties,
their relation to low ionization BALQSOs, 
and to eventually understand the underlying physics. A first step in
this direction is described in the study of Wills et al. (1998, these
proceedings) which includes a few narrow line quasars.


\begin{references}
\reference{} Baldwin, J. A., et al. 1988, \apj, 327, 103
\reference{} Baldwin, J. A., et al. 1996, \apj, 461, 664
\reference{} Boller, T., Brandt, W. N., \& Fink, H. 1996, A\&A, 305, 53
\reference{} Boroson, T. A., \& Meyers, K. A. 1992, \apj, 397, 442
\reference{} Boroson, T. A. \& Green, R. F., 1992, \apjs, 80, 109 
\reference{} Espey, B. R., et al. 1989, \apj, 342, 666
\reference{} Ferland, G. J., et al. 1992, \apj, 387, 95
\reference{} Fiore, F., Laor, A., Elvis, M., Nicastro, F., \& Giallongo, E.
1998, ApJ, 503, 607
\reference{} Gaskell, C. M. 1982, \apj, 263, 79 
\reference{} Hamann, F. 1997, \apjs, 109, 279
\reference{} Hamann, F., Shields, J. C., Ferland, G. J., \& Korista, K. T.
1995, \apj, 454, 688
\reference{} Hartig, G. F. \& Baldwin, J. A. 1986, \apj, 302, 64
\reference{} Johansson, S. \& Jordan, C. 1984, \mnras, 210, 239
\reference{} Korista, K., Baldwin, J., Ferland, G., \& Verner, D.
1997, \apjs, 108, 401
\reference{} Laor, A., et al. 1995, \apjs, 99, 1 
\reference{} Laor, A., et al. 1994, \apj, 435, 611
\reference{} Laor, A., et al. 1997a, \apj, 477, 93
\reference{} Laor, A., Jannuzi, B. T., Green, R. F., \& Boroson, T. A. 1997b,
ApJ, 489, 656 
\reference{} Lipari, S. 1994, \apj, 436, 102
\reference{} Moore, C. E. 1952, An Ultraviolet Multiplet Table, 
NBS Circular 488, (Washington: US Dept. of Commerce)
\reference{} Netzer, H. 1990, in  Active Galactic Nuclei, (Berlin: Springer), 57 
\reference{} Netzer, H., \& Wills, B. J. 1983, \apj, 275, 445 
\reference{} Oke, J. B. \& Lauer, T. R. 1979, \apj, 230, 360
\reference{} Osterbrock, D. E. 1989, Astrophysics of Gaseous Nebulae and Active Galactic Nuclei, 
(California: University Science Books)  
\reference{} Penston, M. V. 1987, \mnras, 229, 1p
\reference{} Phillips, M. M. 1976, \apj, 208, 37
\reference{} Phillips, M. M. 1978, \apjs, 38, 187
\reference{} Sprayberry, D. \& Foltz, C. B. 1992, \apj, 390, 39
\reference{} Thompson, K. L. 1991, \apj, 374, 496
\reference{} Turnshek, D. A., et al. 1994, \apj, 428, 93
\reference{} Tytler, D., \& Fan, X. M. 1992, \apjs, 79, 1 
\reference{} Weymann, R. J., et al. 1991, \apj, 373, 23
\reference{} Wilkes, B. J. 1984, \mnras, 207, 73 
\reference{} Wilkes, B. J. 1986, \mnras, 218, 331 
\reference{} Wills, B. J., Netzer, H., \& Wills, D. 1980, \apj, 242, L1 
\reference{} Wills, B. J., Netzer, H. \& Wills, D. 1985, \apj, 288, 94
\reference{} Wills, B. J., et al. 1995, \apj, 447, 139
\end{references}
\end{document}